\documentclass[
amsmath,
prd,nofootinbib,floatfix,11pt,
]{revtex4}

\usepackage{axodraw2}
\usepackage{graphicx}
\usepackage{setspace}
\usepackage{bm}

\newcommand\beq{\begin{eqnarray}}
\newcommand\eeq{\end{eqnarray}}
\def\lsim{\mathrel{\rlap{\lower4pt\hbox{$\sim$}}
    \raise1pt\hbox{$<$}}}                
\def\gsim{\mathrel{\rlap{\lower4pt\hbox{$\sim$}}
    \raise1pt\hbox{$>$}}}            

\allowdisplaybreaks
\def\intI{{\cal I}}
\newcommand\MSbar{$\overline{\rm{MS}}$ }

\begin{document}

\renewcommand{\theequation}{\arabic{section}.\arabic{equation}}
\renewcommand{\thefigure}{\arabic{section}.\arabic{figure}}
\renewcommand{\thetable}{\arabic{section}.\arabic{table}}

\title{\Large \baselineskip=20pt 
Three-loop QCD corrections to the electroweak boson masses}
\author{Stephen P.~Martin}
\affiliation{
\mbox{\it Department of Physics, Northern Illinois University, DeKalb IL 60115}}

\begin{abstract}\normalsize \baselineskip=15pt 
I find the three-loop corrections at leading order in QCD to the physical masses of the Higgs, $W$, and $Z$ bosons in the Standard Model. The results are obtained as functions of the \MSbar Lagrangian parameters only, using the tadpole-free scheme for the vacuum expectation value. The dependences of the computed masses on the renormalization scale are found to be smaller than present experimental uncertainties in each case. In the case of the Higgs boson mass, the new result is the state-of-the-art, while the results for $W$ and $Z$ are in good numerical agreement with corresponding results in the on-shell and hybrid schemes. These results are now included in the {\tt SMDR} (Standard Model in Dimensional Regularization) computer code. 
\end{abstract}

\maketitle

\vspace{-0.2in}

\tableofcontents

\baselineskip=15.3pt

\section{Introduction \label{sec:intro}}
\setcounter{equation}{0}
\setcounter{figure}{0}
\setcounter{table}{0}
\setcounter{footnote}{1}

Since the discovery of the Higgs boson in 2012, the Standard Model is a mathematically complete theory, for which precision calculations can be performed.  In addition to providing a test of the agreement of the theory with experiment, this allows us to obtain accurate results for the short-distance Lagrangian parameters, suitable for matching to candidate ultraviolet completions. 
The goal of this paper is to report the 3-loop QCD contributions to the pole masses of the $W$, $Z$, and Higgs bosons in the Standard Model. In the case of the Higgs boson mass, the result obtained is the new state-of-the-art result, including the complete set of 2-loop effects as well
as the three-loop terms proportional to $\alpha_S^2 y_t^2$ including all momentum-dependent effects, as well as the three-loop terms proportional to $\alpha_S y_t^4$ and $y_t^6$ in the approximation that $M_h^2 \ll M_t^2$. 

The results below are given in the pure \MSbar renormalization scheme \cite{Bardeen:1978yd,Braaten:1981dv} based on dimensional regularization \cite{Bollini:1972ui,Ashmore:1972uj,Cicuta:1972jf,tHooft:1972fi,tHooft:1973mm}, so that all independent inputs are running Lagrangian parameters. The calculation is also based on the tadpole-free scheme for the Higgs vacuum expectation value (VEV), which is defined to be the minimum of the exact Landau gauge effective potential, currently known in an approximation at full 3-loop order \cite{Ford:1992pn,Martin:2001vx,Martin:2013gka,Martin:2017lqn}
with the leading 4-loop order QCD part \cite{Martin:2015eia} and resummation of the Goldstone boson contributions \cite{Martin:2014bca,Elias-Miro:2014pca}. The tadpole-free VEV scheme has a formally faster convergence in perturbation theory than schemes based on a tree-level VEV definition, since in the latter the tadpole diagrams necessarily introduce inverse powers of the Higgs self-coupling $\lambda$. The price to be paid for this improvement is that the validity of the calculations is restricted to the Landau gauge fixing prescription in the electroweak sector.

Previous 2-loop calculations of the $W$ and $Z$ masses have been given in refs.~\cite{Jegerlehner:2001fb,Jegerlehner:2002em,Degrassi:2014sxa,Kniehl:2015nwa}, using the tree-level definition for the VEV. In addition, there is a long history of calculations of the $\rho$ parameter including up to 4-loop order QCD contributions \cite{vanderBij:1986hy,Djouadi:1987gn,Djouadi:1987di,Kniehl:1989yc,Halzen:1990je,Barbieri:1992nz,Djouadi:1993ss,Fleischer:1993ub,Avdeev:1994db,Chetyrkin:1995ix,Chetyrkin:1995js,Degrassi:1996mg,Freitas:2000gg,vanderBij:2000cg,Freitas:2002ja,Awramik:2002wn,Onishchenko:2002ve,Faisst:2003px,
Awramik:2003ee,Awramik:2003rn,Schroder:2005db,Chetyrkin:2006bj,Boughezal:2006xk}, which can be used to relate the $W$ boson on-shell mass to the $Z$-boson mass. The present paper relies on a quite different organization of perturbation theory, by taking all physical masses as outputs including the $W$ and $Z$ boson pole masses separately, rather than using the $Z$ boson on-shell mass as an input. The complete 2-loop $W$ and $Z$ boson pole squared masses in the scheme adopted in this paper were given in refs.~\cite{Martin:2015lxa} and \cite{Martin:2015rea} respectively. The present paper will add the 3-loop QCD contributions to those results in a consistent way.

In the case of the Higgs boson pole squared mass, ref.~\cite{Bezrukov:2012sa} provided the mixed QCD/electroweak parts, ref.~\cite{Degrassi:2012ry} gave results in the gauge-less limit in which $g,g'$ are neglected in the 2-loop part, and ref.~\cite{Buttazzo:2013uya}
gave an interpolating formula for the full 2-loop approximation in a hybrid \MSbar/on-shell scheme. In ref.~\cite{Martin:2014cxa}, the full 2-loop corrections were extended to include the 3-loop contributions in the gauge-less effective potential limit (formally, $g_3^2, y_t^2 \gg \lambda, g^2, g^{\prime 2}$,
where $g_3,g,g'$ are the gauge couplings, $y_t$ is the top-quark Yukawa coupling, and $\lambda$ is the Higgs self-coupling) using the pure \MSbar tadpole-free scheme. The present paper will extend this further to include the momentum-dependent parts of the leading QCD contribution to the Higgs boson self-energy in the calculation of the pole squared mass. 

To specify notation, the complex pole squared masses for the electroweak bosons are 
each given in the loop-expansion form
\beq
s_{\rm pole}^X \,\equiv\, (M_X - i \Gamma_X/2)^2 
\,=\, m_X^2 + \frac{1}{16\pi^2} \Delta_X^{(1)} + 
 \frac{1}{(16\pi^2)^2} \Delta_X^{(2)} +
 \frac{1}{(16\pi^2)^3} \Delta_X^{(3)} + \ldots
\label{eq:sXpole}
\eeq
with $X = W$, $Z$, and $h$. The complete 2-loop contributions given in refs.~\cite{Martin:2014cxa,Martin:2015lxa,Martin:2015rea} were written in terms of master integrals defined in refs.~\cite{Martin:2003qz,Martin:2005qm}, the latter of which provided a computer program {\tt TSIL} (Two-loop Self-energy Integral Library) for their efficient numerical evaluation. 
The computer program {\tt SMDR} (Standard Model in Dimensional Regularization) \cite{Martin:2019lqd} incorporates these calculations of the $W,Z,$ and Higgs physical masses and many other results within the pure \MSbar tadpole-free scheme, matching observables to Lagrangian parameters. Another public code {\tt mr} \cite{Kniehl:2016enc} provides similar functionality, but using the tree-level VEV scheme.   

For the vector bosons, it is important to note that the standard practice in experimental papers and by the Review of Particle Properties (RPP) \cite{Zyla:2020zbs} from the Particle Data Group (PDG) is to report the on-shell masses found from 
a variable-width Breit-Wigner linewidth fit, which should be related to the complex pole mass and width $M_X$ and $\Gamma_X$ defined in eq.~(\ref{eq:sXpole}) by
\beq
M^{\rm PDG} &=& M \frac{1 + \delta}{\sqrt{1 - \delta}},
\label{eq:MPDG}
\\
\Gamma^{\rm PDG} &=& \Gamma \frac{1 + \delta}{(1 - \delta)^{3/2}},
\label{eq:GammaPDG}
\eeq 
where
\beq
\delta &=& \Gamma^2/4M^2. 
\label{eq:deltaPDGpole}
\eeq
(In this paper, the superscript ``PDG" refers to the convention used by the PDG, and not to the averaged experimental results
produced by the PDG in the RPP.)
To add to the potential for confusion, in refs.~\cite{Martin:2014cxa,Martin:2015lxa,Martin:2015rea} by the present author, and many publications by other authors, a different parameterization for complex pole masses has been used,
denoted here by:
\beq
s_{\rm pole} = M^{\prime 2} - i \Gamma' M',
\eeq
which is related to the $M$ and $\Gamma$ in eq.~(\ref{eq:sXpole}) by
\beq
M' &=& M \sqrt{1 - \delta} 
\label{eq:Mpoleold}
\\
\Gamma' &=& \Gamma/\sqrt{1 - \delta} .
\label{eq:Gammapoleold}
\eeq
The $(M^{\rm PDG},\, \Gamma^{\rm PDG})$ and $(M',\, \Gamma')$ parameterizations can be considered to contain
the same information as $(M,\, \Gamma)$, through the defining relations in eqs.~(\ref{eq:MPDG})-(\ref{eq:deltaPDGpole}) and (\ref{eq:Mpoleold})-(\ref{eq:Gammapoleold}). However, as emphasized in a recent paper \cite{Willenbrock:2022smq},
the $(M,\, \Gamma)$ parameterization defined by eq.~(\ref{eq:sXpole}) has the clear advantage  that  $\Gamma = 1/\tau$ is precisely the inverse mean lifetime of the particle, unlike $\Gamma^{\rm PDG}$ and $\Gamma'$. In the following $s_{\rm pole}$ will be computed, but the information that it contains must be converted to $M^{\rm PDG}$ to compare directly with the results quoted by the PDG and experimental collaborations.
The $W$ and $Z$ PDG-convention masses that are almost always quoted are respectively about 0.020 and 0.026 GeV larger than the pole masses $M_W$ and $M_Z$,
and about 0.027 and 0.034 GeV larger than $M_W'$ and $M_Z'$. The experimental values from the 2021 update of the 2020 RPP are\footnote{After the first version of the present paper, the CDF collaboration released \cite{CDF:2022hxs} a new measurement of
the $W$ mass that is substantially higher, $M_W^{\rm PDG} = 80.4335 \pm 0.0064_{\rm stat} \pm 0.0069_{\rm syst}$ GeV. See Figs.~\ref{fig:MW} and \ref{fig:MWcomparison} below.} $M_{Z}^{\rm PDG} = 91.1876 \pm 0.0021$ GeV and $M_{W}^{\rm PDG} = 80.379 \pm 0.012$ GeV and
$M_h = 125.25 \pm 0.17$ GeV. 
The Higgs boson width (about 4.1 MeV, according to theory) is so small that the numerical distinction between the PDG-convention and complex pole mass versions of the real part $M_h$ is negligible.

The 3-loop integrals to be used below have been defined and discussed 
in sections IV, VI, and VII of
ref.~\cite{Martin:2021pnd}. The master integrals are given there as a renormalized $\epsilon$-finite basis, defined so that expansions of integrals to positive powers in $\epsilon$ will never be needed, even when the results of the present paper are (eventually) extended to 4-loop order or beyond. Denoting the lists of 1-loop, 2-loop, and 3-loop renormalized $\epsilon$-finite master integrals by $\intI_j^{(1)}$, $\intI_j^{(2)}$, and $\intI_j^{(3)}$, respectively, then the general form of a 3-loop contribution to the pole mass of $X = W,Z,$ or $h$ is
\beq
\Delta_X^{(3)}
\, = \,
\sum_j c^{(3)} \intI_j^{(3)} 
+ \sum_{j,k} c^{(2,1)}_{j,k} \intI_j^{(2)} \intI_k^{(1)} 
+ \sum_{j,k,l} c^{(1,1,1)}_{j,k,l} \intI_j^{(1)} \intI_k^{(1)} \intI_l^{(1)} 
&&
\nonumber \\
+ \sum_{j} c^{(2)}_j \intI_j^{(2)} 
+ \sum_{j,k} c^{(1,1)}_{j,k} \intI_j^{(1)} \intI_k^{(1)} 
+ \sum_{j} c^{(1)}_{j} \intI_j^{(1)} 
+ c^{(0)} 
,
&& 
\label{eq:generalform}
\eeq
where all of the coefficients $c^{(3)}, c^{(2,1)}_{j,k}, \ldots ,c^{(0)}$ are 
dimensionless \MSbar couplings multiplied by 
rational functions of the \MSbar top-quark squared mass
\beq
t &=& y_t^2 v^2/2
\eeq
and either $s = W, Z,$ or $h$ as appropriate, where
\beq
W &=& g^2 v^2/4
,
\\
Z &=& (g^2 + g^{\prime 2}) v^2/4
,
\\
h &=& 2 \lambda v^2
.
\eeq
The VEV $v$ is defined to be the minimum of the \MSbar
effective potential in Landau gauge at all orders in perturbation theory, so that
the sum of all Higgs tadpole diagrams vanishes. 
Note that the name of each particle is being used as a synonym for the tree-level \MSbar squared mass in the tadpole-free scheme. (All other fermions are taken to be massless, except in the 1-loop parts $\Delta_X^{(1)}$.) 
Note also that the tree-level \MSbar squared masses $t$, $W$, $Z$, and $h$ are not gauge invariant, but are specific to Landau gauge, due to their dependence on the VEV. However, as is well-known, the complex pole masses [and thus the PDG-convention masses for $W$ and $Z$, defined by eqs.~(\ref{eq:MPDG})-(\ref{eq:deltaPDGpole})] are gauge-invariant.

The loop integrals include logarithmic dependences on the \MSbar renormalization scale $Q$, written in this paper in terms of
\beq
L_t &\equiv& \ln(t/Q^2),
\\
L_{-s} &\equiv& \ln(s/Q^2) - i \pi,
\eeq 
for the external momentum invariant $s$, which has a positive infinitesimal imaginary part. In the 3-loop parts $\Delta_X^{(3)}$, the integrals will always be evaluated at external momentum invariant equal to the tree-level squared mass, $s = W$, $Z$, or $h$. This is just as consistent as choosing to evaluate them at the (real part of) the corresponding pole squared mass instead, as the difference is of 4-loop order, and numerically small.

In order to provide more opportunities for checks, the results below will be
given in terms of $SU(3)_c$ group theory quantities
\beq
C_G = N_c = 3,\qquad C_F = (N_c^2-1)/2 N_c = 4/3,\qquad T_F = 1/2,\qquad n_g = 3.
\label{eq:SMgrouptheory}
\eeq
Here $N_c$ is the number of colors, $C_G$ and $C_F$ are the quadratic Casimir invariants of the adjoint and fundamental representations respectively, $T_F$ is the Dynkin index of the fundamental representation, and $n_g$ is the number of fermion generations.

For numerical results shown below, I will use a benchmark Standard Model designed to give output parameters in agreement with the current central values of the 2021 update of the 2020 RPP \cite{Zyla:2020zbs}:
\beq
&&
M_t \>=\> \mbox{172.5 GeV}
,
\qquad
M_h \>=\> \mbox{125.25 GeV}
,
\qquad
M_{Z}^{\rm PDG} \>=\> \mbox{91.1876 GeV}
,
\nonumber
\\
&&
G_F \>=\> 1.1663787 \times 10^{-5} \> {\rm GeV}^2
,
\qquad
\alpha_0 \>=\> 1/137.035999084
,
\qquad
\alpha_S^{(5)}(M_Z) = 0.1179
,
\nonumber
\\
&&
m_b(m_b) \>=\> \mbox{4.18 GeV}
,
\qquad
m_c(m_c) \>=\> \mbox{1.27 GeV}
,
\qquad
m_s(\mbox{2 GeV}) \>=\> \mbox{0.093 GeV}
,
\nonumber
\\
&&
m_d(\mbox{2 GeV}) \>=\> \mbox{0.00467 GeV}
,
\qquad
 m_u(\mbox{2 GeV}) \>=\> \mbox{0.00216 GeV}
 ,
\qquad
M_\tau \>=\> \mbox{1.77686 GeV}
,
\nonumber
\\
&&
M_\mu \>=\> \mbox{0.1056583745 GeV}
,
\qquad
M_e \>=\> \mbox{0.000510998946 GeV}
,
\nonumber
\\
&&
\Delta \alpha^{(5)}_{\rm had}(M_Z) \>=\> 0.02766
.
\label{eq:referencemodelOS}
\eeq
Using the latest version 1.2 of the computer program {\tt SMDR} \cite{Martin:2019lqd}, which incorporates the new results of the present paper, these are best fit by the \MSbar input parameters (using the tadpole-free scheme for the Landau-gauge VEV, and writing $g_3$ for the QCD coupling in the full 6-quark Standard Model theory):
\beq
Q_0 &=& 172.5\>{\rm GeV},
\nonumber
\\
v(Q_0) &=&         
                    246.603216913\>{\rm GeV},
\qquad
\lambda(Q_0) \>=\> 
                    0.12639276585,
\nonumber
\\
g_3(Q_0) &=&       
                    1.16300624875,\qquad\>\>\,
g(Q_0) \>=\>       
                   0.647606757306,\qquad\>                   
g'(Q_0) \>=\>      
                    0.358550211695,
\nonumber
\\
y_t(Q_0) &=&       
                    0.93157701535,\qquad
y_b(Q_0) \>=\>     
                    0.015503239387,\qquad
y_\tau(Q_0) \>=\>  
                    0.0099944376213,
\nonumber
\\
y_c(Q_0) &=&       
                    0.003394710569,
\qquad
y_s(Q_0) \>=\>     
                    0.0002916507520,
\qquad
y_\mu(Q_0)  \>=\>  
                    0.0005883797990,\phantom{xxx}
\nonumber
\\
y_d(Q_0) &=&       
                    1.464523924362\times 10^{-5},
\qquad
y_u(Q_0) \>=\>     
                    6.739112138367\times 10^{-6},
\nonumber
\\
y_e(Q_0) &=&       
                    2.792980305214\times 10^{-6}.
\label{eq:referencemodelMSbar}
\eeq
Here I have included many more significant digits than justified by the theoretical errors, merely for the sake of reproducibility. These \MSbar quantities can be run to a different renormalization scale choice $Q$, where the pole squared masses can be recomputed. In the idealized case, the pole squared masses, being observables, would be independent of the scale $Q$ at which they are computed.

The renormalization group running is carried out using the state-of-the-art beta functions for the Standard Model. 
The 2-loop and 3-loop beta functions were found in 
\cite{MVI,MVII,Jack:1984vj,MVIII,Luo:2002ey} and
\cite{Tarasov,Mihaila:2012fm,Chetyrkin:2012rz,Bednyakov:2012rb,Bednyakov:2012en,
Chetyrkin:2013wya,Bednyakov:2013eba,Bednyakov:2013cpa,Bednyakov:2014pia}, respectively.
The 4-loop beta function for the QCD coupling $g_3$ was found in \cite{vanRitbergen:1997va,Czakon:2004bu,
Bednyakov:2015ooa,Zoller:2015tha,Poole:2019txl}
in the approximation that only $g_3$, $y_t$, and $\lambda$ are included. The pure QCD 5-loop beta functions were obtained in \cite{Baikov:2016tgj,Herzog:2017ohr}, and the 
4-loop and 5-loop QCD contributions to the quark Yukawa beta functions were obtained
in refs.~\cite{Chetyrkin:1997dh,Vermaseren:1997fq} and ref.~\cite{Baikov:2014qja} respectively, and the 4-loop QCD contributions to the beta function of the Higgs self-coupling $\lambda$ were obtained from \cite{Martin:2015eia,Chetyrkin:2016ruf}. Finally,
the complete 4-loop beta functions for the three gauge couplings have been provided by
\cite{Davies:2019onf}. All of these results have been included in the latest version of
the code {\tt SMDR}, which was used to carry out the numerical computations described below.
The code also implement results for multi-loop threshold matching of electroweak couplings \cite{Fanchiotti:1992tu,Erler:1998sy,Degrassi:2003rw,Degrassi:2014sxa,Kniehl:2015nwa,Martin:2018yow},
the QCD coupling \cite{Larin:1994va,Chetyrkin:1997un,Grozin:2011nk,Schroder:2005hy,Chetyrkin:2005ia,Bednyakov:2014fua},
and quark and lepton masses \cite{Tarrach:1980up,Gray:1990yh,Melnikov:2000qh,Chetyrkin:2000yt,Kniehl:2004hfa,Schmidt:2012az,Kniehl:2014yia,Marquard:2015qpa,Liu:2015fxa,Marquard:2016dcn,Bednyakov:2016onn,Herren:2017osy}.

\section{The $Z$ boson pole mass\label{sec:Zboson}}
\setcounter{equation}{0}
\setcounter{figure}{0}
\setcounter{table}{0}
\setcounter{footnote}{1}

Consider the $Z$-boson complex pole squared mass, $s_{\rm pole}^Z$, in the form of eq.~(\ref{eq:sXpole}).
The complete 1-loop and 2-loop contributions $\Delta_Z^{(1)}$ and $\Delta_Z^{(2)}$ were given in the tadpole-free pure \MSbar scheme in ref.~\cite{Martin:2015rea}.
The 3-loop QCD part can be split into contributions from 13 distinct classes of self-energy diagrams with different group theory structures, using the quantities defined in eq.~(\ref{eq:SMgrouptheory}):
\beq
\Delta_Z^{(3),g_3^4} &=&
 g_3^4 N_c C_F \biggl \{
(a_{u_L}^2 + a_{u_R}^2) \left [C_G \Delta_Z^{(3,a)} + C_F \Delta_Z^{(3,b)} + 
T_F \Delta_Z^{(3,c)} + (2 n_g-1) T_F \Delta_Z^{(3,d)} \right ]
\nonumber \\ && 
+ 
2 a_{u_L} a_{u_R} \left [C_G \Delta_Z^{(3,e)} + C_F \Delta_Z^{(3,f)} + 
T_F \Delta_Z^{(3,g)} + (2 n_g-1) T_F \Delta_Z^{(3,h)} \right ]
\nonumber \\ && 
+ (g^2 + g^{\prime 2}) T_F \Delta_Z^{(3,i)}
+ \left [(n_g-1) (a_{u_L}^2 + a_{u_R}^2) + n_g (a_{d_L}^2 + a_{d_R}^2) \right ]
\Bigl [C_G \Delta_Z^{(3,j)} 
\nonumber \\ && 
+ C_F \Delta_Z^{(3,k)} 
+ T_F \Delta_Z^{(3,l)} 
+ (2 n_g-1) T_F \Delta_Z^{(3,m)} \Bigr ]
\biggr \}
,
\phantom{xxx}
\label{eq:DeltaZ3grouptheory}
\eeq
where the tree-level couplings of the $Z$ boson to up-type and down-type quarks are
\beq
a_{u_R} &=& -\frac{2}{3} \frac{g^{\prime 2}}{\sqrt{g^2 + g^{\prime 2}}},
\qquad\quad
a_{u_L} \>=\> \frac{1}{2} \sqrt{g^2 + g^{\prime 2}} + a_{u_R},
\\
a_{d_R} &=& \frac{1}{3} \frac{g^{\prime 2}}{\sqrt{g^2 + g^{\prime 2}}},
\qquad\quad\>\>
a_{d_L} \>=\> -\frac{1}{2} \sqrt{g^2 + g^{\prime 2}} + a_{d_R}
.
\eeq

\begin{figure}[!t]
\begin{center}
\includegraphics[width=11cm,angle=0]{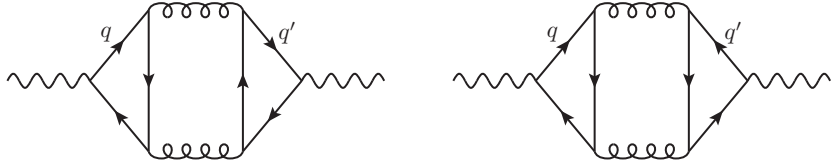}
\caption{\label{fig:Zsinglet}Three-loop contribution to the $Z$ boson mass from diagrams involving two triangle quark loops, which give a non-vanishing contribution with a consistent treatment of the axial vector coupling. These contributions are individually divergent for each of $(q,q') = (t,t), (t,b), (b,t), (b,b)$, but are finite and gauge invariant after the combination. Contributions involving sums over other $(q,q')$ quark doublet combinations vanish in the massless quark limit.}
\end{center}
\end{figure}

Most of the three-loop diagrams are straightforward to set up, and can be carried out with a naive treatment of $\gamma_5$, taken to anti-commute with all of the other gamma matrices. The known exception to this
is the double triangle diagrams shown in Figure \ref{fig:Zsinglet}, which feature two distinct triangle quark loops each containing a $\gamma_5$ from the axial vector coupling to the $Z$ boson.
(The vector couplings to the $Z$ boson give vanishing contributions for the sum of these diagrams.) 
The contributions from $(q,q') = (t,t), (t,b), (b,t), (b,b)$ are separately divergent, but their sum is finite and gauge invariant. Therefore, for these diagrams only, one can use the prescription
\cite{Akyeampong:1973xi,Larin:1993tq} 
\beq
\gamma^\mu \gamma_5 \rightarrow \frac{i}{6} 
\epsilon^{\mu\nu\rho\sigma} \gamma_{\nu} \gamma_{\rho} \gamma_{\sigma},
\eeq 
based on the 't Hooft-Veltman treatment \cite{tHooft:1972fi} of $\gamma_5$,
and then carry out the Lorentz algebra in 4 dimensions before reducing to master integrals in $d$ dimensions. The result is  
the contribution $\Delta_Z^{(3,i)}$ in eq.~(\ref{eq:DeltaZ3grouptheory}). The contributions from diagrams with one or both of $q$ and $q'$ summed over
the other quark doublets $(u,d)$ and $(c,s)$ vanish, because
the axial couplings $a_{q_L} - a_{q_R}$ for down-type and up-type quarks have the same magnitude and opposite sign, and they are being treated as mass degenerate (specifically, massless). 
The result for general non-zero $s = Z$ found here reduces to $\Delta_Z^{(3,i)} \rightarrow 21 \zeta_3$ for $s = 0$, which agrees with the original calculation in that limit \cite{Anselm:1993uq} and with the corresponding contribution to the $\rho$ parameter obtained in \cite{Avdeev:1994db,Chetyrkin:1995ix,Schroder:2005db}.

The contributions from the diagrams in which the $Z$ boson couples directly to a single massless (in the present approximation, non-top) quark loop are relatively simple, and can be written as:
\beq
\Delta_Z^{(3,j)} &=& Z \left (
-\frac{44215}{324} + \frac{908}{9} \zeta_3 + \frac{40}{3} \zeta_5 
+  \left [41 - \frac{88}{3}\zeta_3 \right ] L_{-Z}  - \frac{11}{3} L_{-Z}^2 
\right )
,
\label{eq:DeltaZ3j}
\\
\Delta_Z^{(3,k)} &=& Z \left (\frac{143}{9}  + \frac{148}{3} \zeta_3 - 80 \zeta_5 - L_{-Z}
\right )
,
\\
\Delta_Z^{(3,l)} &=&
\frac{16}{27}Z (7t + 3Z) I_{7c}(0, 0, 0, 0, 0, t, t)
+\frac{16}{243} (128 t + 43 Z) I_{6c} (0, 0, 0, 0, t, t)
\nonumber \\ &&
- \frac{8}{27} (18 t + 7 Z) I_{6f} (0, 0, 0, 0, t, t)
+ \frac{32}{243 t} (5 Z - 17 t) I_{5c} (0, 0, 0, t, t)
+ \frac{160}{81 t} I_4 (0,0,t,t)
\nonumber \\ &&
+ \left (\frac{896}{243}\zeta_3 -\frac{8276}{729}\right ) t
+ \frac{2599}{243}{Z} + \frac{80}{2187}\frac{Z^2}{t} 
+ \left (\frac{11144}{243} t - \frac{224}{9} \zeta_3 t - \frac{3320}{243} Z\right ) L_{t}
\nonumber \\ &&
-\frac{352}{81} t L_t^2
-\frac{112}{243} t L_t^3
+ \frac{4}{243} (884 t - 217 Z) L_{-Z}
+ \left ( \frac{160}{27} Z - \frac{32}{3} t \right ) L_{t} L_{-Z}
\nonumber \\ &&
+ \frac{272}{81} t L_t^2 L_{-Z}
+ \frac{20}{81} Z L_{-Z}^2
- \frac{16}{243} (17 t + 10 Z) L_{t} L_{-Z}^2
,
\label{eq:DeltaZ3l}
\\[5pt]
\Delta_Z^{(3,m)} &=& Z \left ( \frac{3701}{81} - \frac{304}{9} \zeta_3 
+  \left [\frac{32}{3} \zeta_3 -\frac{44}{3} \right ] L_{-Z} + \frac{4}{3} L_{-Z}^2
\right )
.
\label{eq:DeltaZ3m}
\eeq
Here, $\Delta_Z^{(3,l)}$ contains a top-quark loop that corrects a gluon propagator, rather than connecting to the external $Z$ boson.
The remaining contributions in eq.~(\ref{eq:DeltaZ3grouptheory}) are much more complicated, and are given in an ancillary file {\tt DeltaZ3} provided with this paper.
Each of the contributions has the form of eq.~(\ref{eq:generalform}), with master integrals chosen in ref.~\cite{Martin:2021pnd}:
\beq
\intI^{(1)} &=& \{\hspace{0.4pt} A(t),\> B(0,0),\> B(t,t)\, \}
,
\label{eq:I1forZ}
\\
\intI^{(2)} &=& \{\hspace{0.4pt} \zeta_3,\, V(t, t, 0, t),\, M(t, t, t, t, 0),\, M(0, t, 0, t, t) \, \}
,
\\
\intI^{(3)} &=& \{\hspace{0.4pt}
\zeta_5,\,
H(0, 0, t, 0, t, t),\,
H(0, t, t, t, 0, t),\,
I_{4}(t, t, t, t),\, 
I_{5a}(t, 0, t, 0, t),\,
I_{5b}(0, t, t, t, t), 
\nonumber \\ &&
I_{5c}(t, t, t, t, t),\, 
I_{6c}(t, t, t, 0, t, t),\,
I_{6c2}(t, t, t, 0, 0, 0),\, 
I_{6d}(0, t, t, t, t, 0),\, 
I_{6d}(t, 0, t, 0, t, 0),
\nonumber \\ &&
I_{6d}(t, 0, t, t, 0, t),\, 
I_{6e}(0, 0, 0, 0, t, t),\, 
I_{6e}(0, t, t, t, 0, t),\, 
I_{6e}(t, t, t, 0, t, t),\,
I_{6f}(0, 0, 0, 0, t, t),\, 
\phantom{x}
\nonumber \\ &&
I_{6f5}(0, 0, 0, 0, t, t),\, 
I_{7a}(0, 0, t, t, t, t, t),\,
I_{7a}(t, t, t, t, t, t, 0),\, 
I_{7a3}(t, t, t, t, t, t, 0),\,
\nonumber \\ &&
I_{7b}(0, t, t, t, t, 0, 0),\, 
I_{7b}(t, 0, t, t, t, t, 0),\,
I_{7b4}(t, 0, t, t, t, t, 0),\, 
I_{7b4}(t, t, 0, t, t, 0, t),\,
\nonumber \\ &&
I_{7c}(t, t, t, t, 0, 0, 0),\, 
I_{7d}(t, t, 0, t, 0, t, 0),\,
I_{7d}(t, t, 0, t, t, 0, t),\, 
I_{7e}(0, 0, 0, 0, 0, t, t),\,
\nonumber \\ &&
I_{7e}(0, 0, t, t, t, 0, 0),\,
I_{8a}(t, 0, t, t, t, t, t, 0),\,
I_{8a}(t, t, t, t, t, 0, 0, 0),\, 
I_{8b}(t, t, t, t, t, 0, 0, t),\,
\nonumber \\ &&
I_{8c}(t, 0, t, t, t, t, t, 0),\,
I_{8c}^{pk}(t, t, t, t, t, 0, 0, t)
\,\}
,
\label{eq:I3forZ}
\eeq
with $A(t) = t (L_t-1)$ and $B(0,0) = 2 - L_{-Z}$.
However, in eq.~(\ref{eq:DeltaZ3l}) above, I have chosen to write the expression for $\Delta_Z^{(3,l)}$ in terms of candidate master integrals that were solved for in ref.~\cite{Martin:2021pnd}, rather than the master integrals listed above (which are a subset of the ones listed in eq.~(7.4) in ref.~\cite{Martin:2021pnd}, joined by $B(0,0)$ and $\zeta_3$ and $\zeta_5$ from the integrals with all propagators massless). This simplifies the expression somewhat, because the integrals used in eq.~(\ref{eq:DeltaZ3l}) have the same propagator structures as descendants of the underlying Feynman diagrams for the $\Delta_Z^{(3,l)}$ contribution.

As a check of eq.~(\ref{eq:DeltaZ3grouptheory}), I have verified that the full expression for the observable $s_{\rm pole}^Z$ is renormalization group invariant through 3-loop terms proportional to $g_3^4$, using the
derivatives of the master integrals with respect to $Q$ found in the ancillary file
{\tt QddQ} of ref.~\cite{Martin:2021pnd}.

For practical numerical evaluation, after using the Standard Model group theory values in
eq.~(\ref{eq:SMgrouptheory}), and applying the expansions for the master integrals in the ancillary file {\tt Ievenseries} of ref.~\cite{Martin:2021pnd}, I find:
\beq
\Delta_Z^{(3),g_3^4} &=&
 g_3^4 t \left \{(g^2 + g^{\prime 2}) (\delta_1^Z + \delta_2^Z) + a_{u_L} a_{u_R} \delta_3^Z + \left [2 (a_{u_L}^2 + a_{u_R}^2) + 3 (a_{d_L}^2 + a_{d_R}^2) \right ] \delta_4^Z \right \}
 ,
 \phantom{xxx}
 \label{eq:DeltaZ3QCD}
 \eeq
where the series expansions of $\delta_1^Z$, $\delta_2^Z$, $\delta_3^Z$, and $\delta_4^Z$ are given in the ancillary file {\tt DeltaZ3series} to order $r_Z^{18}$, where
\beq
r_Z &\equiv& 
\frac{Z}{4t} \>=\> \frac{g^2 + g^{\prime 2}}{8 y_t^2}.
\eeq 
The contribution $\delta_1^Z$ isolates the results form the double triangle diagrams in Figure \ref{fig:Zsinglet}.
The series expansion coefficients are given both numerically, and analytically in terms of 
$L_t$, $L_{-Z}$, and the constants $\zeta_3$, $\zeta_5$, and 
\beq
c_H' 
\>=\> 
32 {\rm Li}_4(1/2) - 22 \zeta_4 + \frac{4}{3} \ln^2(2) [\ln^2(2) - \pi^2)] 
\>\approx\> 
-13.2665092775\ldots .
\label{eq:definecHp}
\eeq
The series converge for all $r_Z<1$, which is clearly satisfied in actuality. The
first few terms in the expansions are
\beq
\delta_1^Z &=& 50.486 
+ r_Z \left [79.645 + 49.333 (L_t - L_{-Z}) + 8 (L_t - L_{-Z})^2 \right ] 
\nonumber \\ && 
+ r_Z^2 \left [-15.758 + 5.531 (L_t - L_{-Z}) \right ]
+ r_Z^3 \left [-3.066 - 1.493 (L_t - L_{-Z}) \right ]
+ \ldots
\\
\delta_2^Z &=&   
9.978 + 49.258 L_{t} + 18 L_t^2 -30 L_t^3
+ r_Z \left (-113.200 - 90.222 L_{t} + 28 L_t^2 \right )
\nonumber \\ && 
+ r_Z^2 \left (-42.485 - 63.002 L_{t} - 4.8 L_t^2 \right ) 
\nonumber \\ && 
+ r_Z^3 \left (-45.813 - 74.011 L_{t} - 32.914 L_t^2 \right )
+ \ldots
,
\\
\delta_3^Z &=&  
r_Z \left (-687.728 - 298.667 L_{t} + 224 L_t^2 \right ) 
+ r_Z^2 \left (-733.683 - 685.827 L_{t} - 51.200 L_t^2 \right ) 
\nonumber \\ && 
+ r_Z^3  \left (-707.875 - 962.072 L_{t} - 394.971 L_t^2 \right )
+ \ldots
,
\\
\delta_4^Z &=& 
r_Z \left ( -56.799 - 14.758 L_{t}  - 10.667 L_t^2 + 180.381 L_{-Z}
+ 21.333 L_{t} L_{-Z}  - 122.667 L_{-Z}^2  \right )
\nonumber \\ &&
+ r_Z^2 \left [ -88.570 - 33.375 (L_{t} - L_{-Z}) - 3.793 (L_{t} - L_{-Z})^2 \right ]
\nonumber \\ &&
+ r_Z^3 \left [ 4.074 + 2.521 (L_{t} - L_{-Z}) + 0.406 (L_{t} - L_{-Z})^2 \right ] + \ldots
,
\label{eq:delta4Z}
\eeq
It is interesting to note that in the expansion in small $r_Z$, the sub-leading contribution is
numerically comparable to (or even larger than, for smaller $Q$) the leading contribution obtained by $r_Z=0$. This is due mostly to the term proportional to $r_Z L_{-Z}^2$ in the contribution eq.~(\ref{eq:delta4Z})
from massless quark loops, because of the large magnitude of the coefficient $-122.667$ and because
$L_{-Z}^2 = [-i\pi + \ln(Z/Q^2)]^2$ provides up to an order of magnitude enhancement.

\begin{figure}[!t]
  \begin{minipage}[]{0.495\linewidth}
    \includegraphics[width=8.0cm,angle=0]{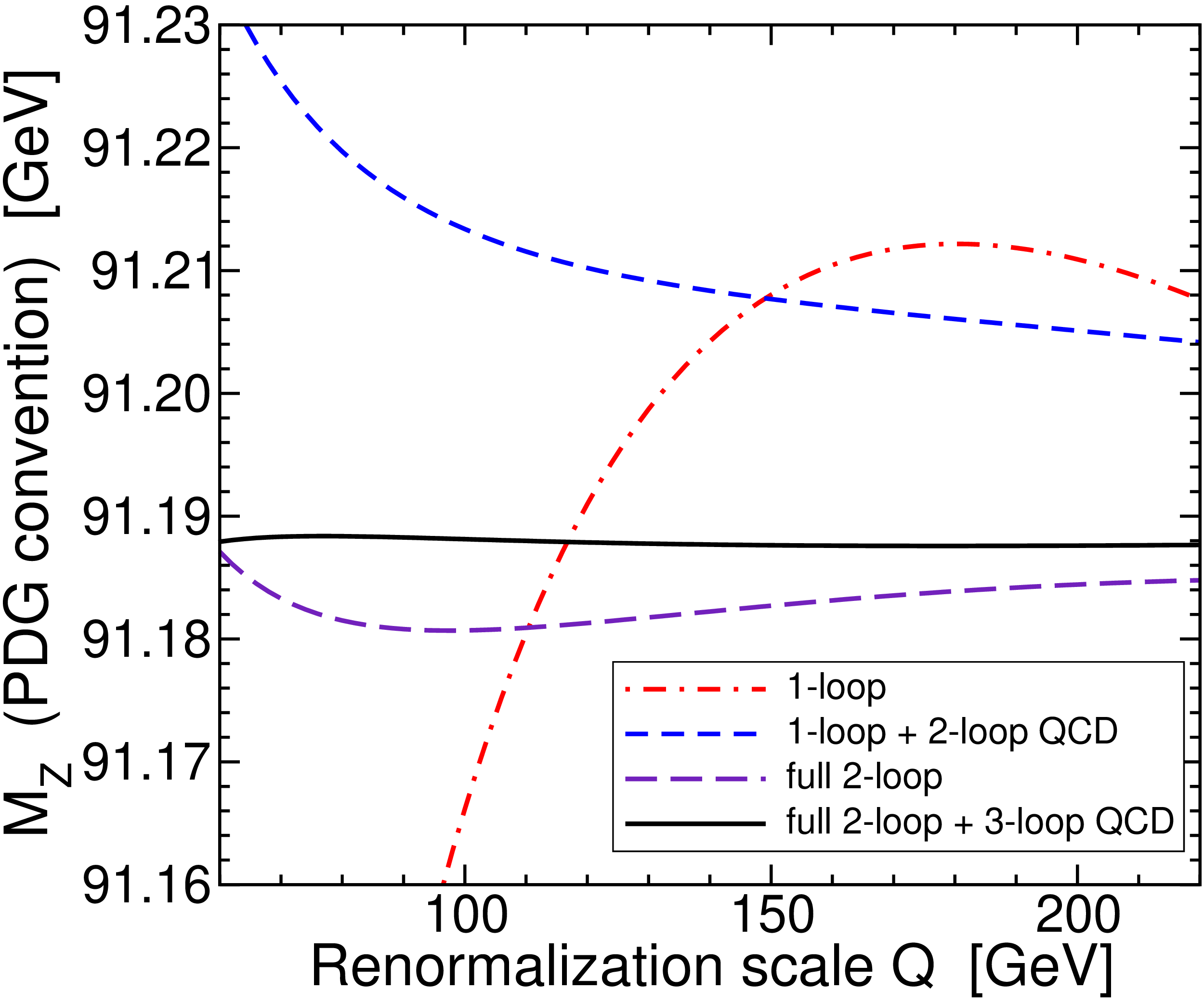}
  \end{minipage}
  \begin{minipage}[]{0.495\linewidth}
    \includegraphics[width=8.0cm,angle=0]{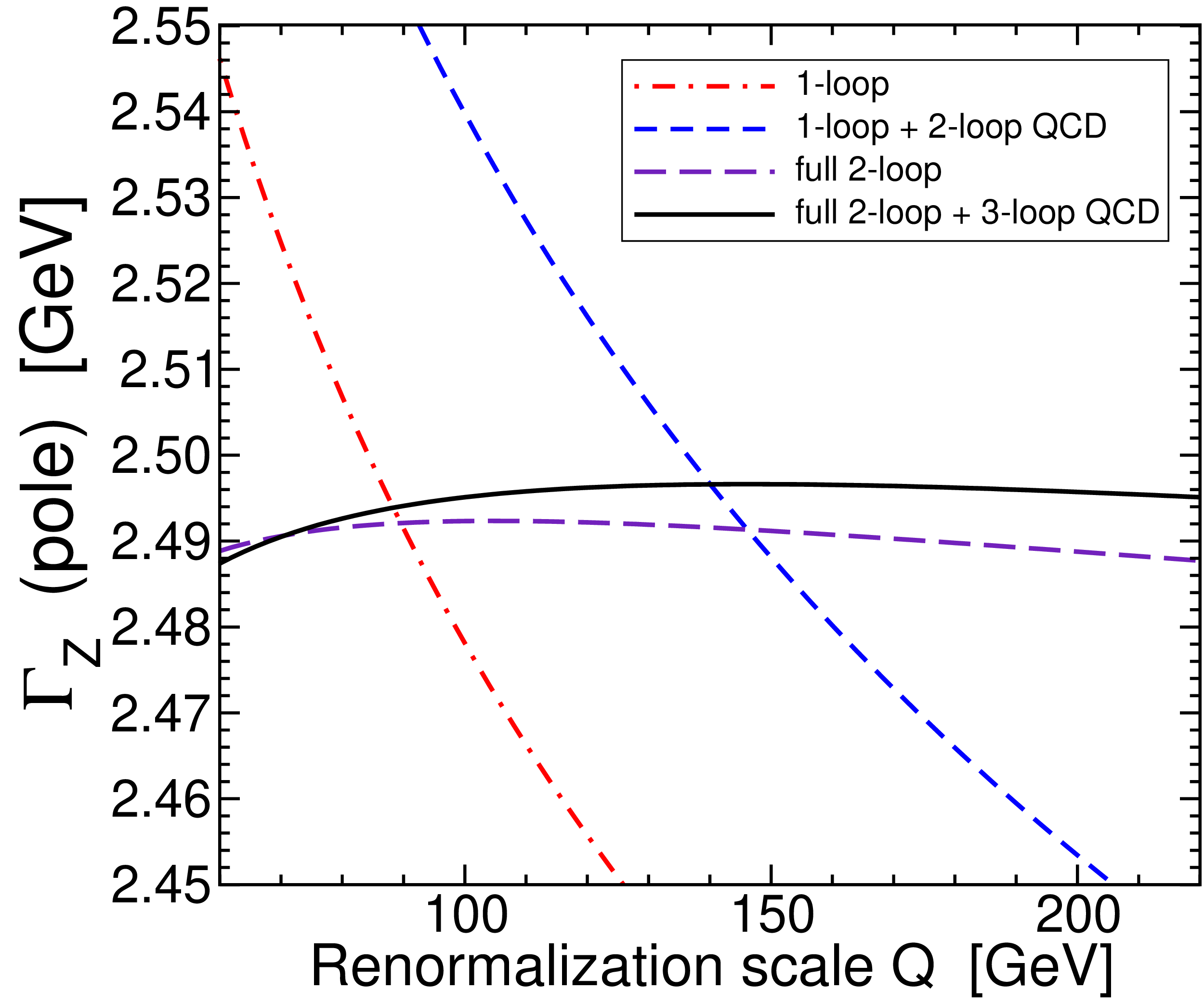}
  \end{minipage}
  \begin{center}
    \begin{minipage}[]{0.95\linewidth}
\caption{\label{fig:MZ}The $Z$ boson mass in the PDG convention $M_Z^{\rm PDG}$ (left panel) and the width $\Gamma_Z$ (right panel), obtained from the calculated complex pole mass $s^Z_{\rm pole} $, as a function of the renormalization scale $Q$. The different lines show various approximations as labeled. The \MSbar input parameters are as given in eq.~(\ref{eq:referencemodelMSbar}), which provide for $M_Z^{\rm PDG} = 91.1876$ GeV when calculated at the renormalization scale $Q=160$ GeV using the full 2-loop plus 3-loop QCD approximation.}
    \end{minipage}
  \end{center}
\end{figure}

The resulting contribution of eq.~(\ref{eq:DeltaZ3QCD}) has now been included in the latest version 1.2 of the code {\tt SMDR} \cite{Martin:2019lqd}.
Figure \ref{fig:MZ} shows the results for the PDG-convention mass $M_Z^{\rm PDG}$ and the width $\Gamma_Z$ obtained from the pole mass, for the \MSbar input parameters given in eq.~(\ref{eq:referencemodelMSbar}).
These benchmark parameters were chosen so that the calculated $M_Z^{\rm PDG}$,
with all known contributions included and using the renormalization scale $Q = 160$ GeV, is equal to the experimental central value 91.1876 GeV. To obtain the results in the figure, the \MSbar input parameters are run to other \MSbar scales $Q$ using the most complete available renormalization group equations (as listed at the end of the Introduction), and $s_{\rm pole}^Z$ is then re-calculated. In the idealized case, the results should not depend on $Q$. I find that with inclusion of the 3-loop QCD corrections, the scale dependence of $M_Z$ is remarkably small, less than 0.8 MeV as $Q$ is varied between 50 GeV and 220 GeV. However, given the larger scale dependence found in section \ref{sec:Wboson} for the similar case of the $W$ boson mass, I surmise that this very mild scale dependence is partly accidental, and the actual theoretical error due to neglecting higher order contributions is likely to be larger.

The scale dependence of $\Gamma_Z$ shown in the right panel of Figure \ref{fig:MZ} is less mild, and not so much improved over the complete 2-loop result, as it varies by a total of about 4 MeV (between minimum and maximum) as $Q$ is varied between 80 GeV and 220 GeV. Note that this determination of $\Gamma_Z$ from the complex pole mass (in which the leading contribution arises only as a 1-loop effect) is essentially one loop order less accurate than a direction calculation of the $Z$-boson decay width (in which the leading contribution is a tree-level effect).
 
\section{The Higgs boson pole mass\label{sec:Hboson}}
\setcounter{equation}{0}
\setcounter{figure}{0}
\setcounter{table}{0}
\setcounter{footnote}{1}

Next, consider the complex pole mass $s^h_{\rm pole}$ for the Standard Model Higgs boson, written in the form of eq.~(\ref{eq:sXpole}). In this section, I extend the results of ref.~\cite{Martin:2014cxa} to include the momentum-dependent 3-loop self-energy corrections to $\Delta_h^{(3)}$ that are proportional to $g_3^4 y_t^2 t$.  Also included below are the 3-loop contributions proportional to $g_3^2 y_t^4 t$ and $y_t^6 t$, in an effective potential approximation, which amounts to  
$g_3^2, y_t^2 \gg \lambda, g^2, g^{\prime 2}$. 
For the $y_t^6 t$ part, I provide below an improvement over the result in \cite{Martin:2014cxa}. Together with the full 2-loop results, these constitute the most complete calculation of the Standard Model Higgs boson mass that is presently available.

The functions $\Delta_h^{(1)}$ and the QCD part of $\Delta_h^{(2)}$ were given in eqs.~(2.46) and (2.47) in ref.~\cite{Martin:2014cxa}, and are evaluated at $s={\rm Re}[s^h_{\rm pole}]$, determined by iteration. The remaining, non-QCD part of $\Delta_h^{(2)}$ was given in an ancillary file of ref.~\cite{Martin:2014cxa},
where the master integrals were also evaluated at $s={\rm Re}[s^h_{\rm pole}]$. However, in the present paper, I adopt a slightly different organization by evaluating the non-QCD part of $\Delta_h^{(2)}$ as exactly the same function but evaluated instead at $s=h$, which is consistent up to 3-loop terms of order $y_t^6 t$. This allows an easier extension to 3-loop order, as indicated below.

For the leading QCD part of $\Delta_h^{(3)}$ proportional to $g_3^4 y_t^2 t$, the 
new result can be written in terms of the contributions of four distinct classes
of self-energy diagrams characterized by their group theory structures:
\beq
\Delta_h^{(3),g_3^4 y_t^2 t} &=& g_3^4 y_t^2 N_c C_F 
\left (
C_G \Delta_h^{(3,a)}
+ C_F \Delta_h^{(3,b)}
+ T_F \Delta_h^{(3,c)}
+ (2 n_g - 1) T_F \Delta_h^{(3,d)}
\right )
.
\eeq
The results for $\Delta_h^{(3,a)}$, $\Delta_h^{(3,b)}$, $\Delta_h^{(3,c)}$, and
$\Delta_h^{(3,d)}$ are somewhat lengthy, and so are given in the ancillary file {\tt DeltaH3} provided with this paper.
They are written in terms of the same list of 
3-loop self-energy master integrals as for the $Z$ boson, listed in 
eqs.~(\ref{eq:I1forZ})-(\ref{eq:I3forZ}), with the exceptions that $I_{8b}(t, 0, t, t, t, t, t, 0)$
is also needed in $\intI^{(3)}$, and $\zeta_5$, $I_{6e}(0, 0, 0, 0, t, t)$,
$I_{6f}(0, 0, 0, 0, t, t)$, $I_{6f5}(0, 0, 0, 0, t, t)$,  $I_{7e}(0, 0, 0, 0, 0, t, t)$,
and $I_{7e}(0, 0, t, t, t, 0, 0)$ are not needed, and of course one should use $s=h$ rather than $s=Z$.

Using the expansions of the master integrals
given in ref.~\cite{Martin:2021pnd}, and setting $s=h$ in 
$\Delta_h^{(3),g_3^4 y_t^2 t}$ (which is consistent up to terms of 4-loop order), and plugging in the group theory constants from eq.~(\ref{eq:SMgrouptheory}), the result becomes a power series in 
\beq
r_h \equiv \frac{h}{4t} = \frac{\lambda}{y_t^2}
,
\eeq
with coefficients that depend on $L_t \equiv \ln(t/Q^2)$ and $L_{-h} \equiv \ln(h/Q^2) - i \pi$
and the constants $\zeta_3$ and $c_H'$ from eq.~(\ref{eq:definecHp}).
The expansion converges for $r_h < 1$, and does so rapidly for the value realized in the Standard Model. It is given to order $r_h^{24}$ in the ancillary file {\tt DeltaH3series},
both in analytic and numerical forms.
The first few terms of the numerical form are: 
\beq
\Delta_h^{(3),g_3^4 y_t^2 t} &=& g_3^4 y_t^2 t \biggl (
 248.122 + 839.197\, L_{t} + 160\,L_t^2 - 736\, L_t^3 
 \nonumber \\ &&
 + r_h \left [ 
 -716.898 - 1546.064\, L_{t} + 336\, L_t^2 + 240\, L_t^3  
 \right ]      
 \nonumber \\ &&
 + r_h^2 \left [ 
 479.663 + 72.770\, L_{t} + 28.444\,L_{-h}
 \right ] 
 \nonumber \\ &&
+ r_h^3 \left [ -27.675 - 83.837\, L_{t} - 5.486\, L_t^2 + 13.274\, L_{-h}  \right ]
+ \ldots
\biggr ) 
.
\eeq 
As a non-trivial check, the result obtained with $r_h=0$ agrees with that provided in the first line of eq.~(3.3) of ref.~\cite{Martin:2014cxa}.
The terms with positive powers of $r_h$ are new in the present paper.

For the part of $\Delta_h^{(3)}$ proportional to $g_3^2 y_t^4 t$, the effective potential approximation gives the second line of eq.~(3.3) of ref.~\cite{Martin:2014cxa}, which is
not improved on in the present paper, but is reproduced here for reference and comparison:
\beq
\Delta_h^{(3),g_3^2 y_t^4 t} &=& g_3^2 y_t^4 t \left (
2764.365 + 1283.716\, L_{t} - 360 \, L_t^2 + 240 \, L_t^3
\right )
.
\eeq
It is interesting that $\Delta_h^{(3),g_3^4 y_t^2 t}$
is numerically smaller than $\Delta_h^{(3),g_3^2 y_t^4 t}$,
despite the parametric relative enhancement $N_c g_3^2/y_t^2$ of the former. In the approximation $r_h=0$, this effect was noted in 
refs.~\cite{Martin:2014cxa,Martin:2013gka} (see the discussion involving eqs.~(6.21)-(6.28) of the former reference) as the result of an unexplained but 
dramatic near-cancellation, and is found here to be not changed by the inclusion of terms higher order in $r_h$.

Finally, for the part of $\Delta_h^{(3)}$ proportional to $y_t^6 t$, the effective potential approximation of ref.~\cite{Martin:2014cxa} can be improved on slightly as follows. In the present paper,
the non-QCD part of $\Delta_h^{(2)}$ is evaluated using master integrals with
external momentum invariant $h$ rather than ${\rm Re}[s^h_{\rm pole}]$. 
Then, due to the fortunate circumstance that the leading 1-loop behavior of $s^h_{\rm pole}-h$ in the limit 
$y_t^2 \gg \lambda, g^2, g^{\prime 2}$ is proportional to $L_t$:
\beq
s^h_{\rm pole}-h &=& \frac{1}{16\pi^2} 4 N_c y_t^2 t L_t,
\eeq
we can fully repair the error in the 3-loop part (caused by using $h$ rather than ${\rm Re}[s^h_{\rm pole}]$ in the 2-loop part), simply by requiring renormalization group invariance of the pole mass. This allows inference of the complete dependence proportional to $y_t^6 t L_t$, due to the explicit dependence on $Q$. 
By demanding (and checking) renormalization group invariance of $s_{\rm pole}^h$ through terms of 3-loop order in the approximation 
$g_3^2, y_t^2 \gg \lambda, g^2, g^{\prime 2}$, 
I find that the end result for the leading non-QCD 3-loop contribution is that eq.~(3.4) of ref.~\cite{Martin:2014cxa} should be replaced by:
\beq
\Delta_{M_h^2}^{(3), y_t^6 t} &=&
y_t^6 t \Bigl [
-3433.724
- 2426.808 L_t
- 101.016 L_t^2
- 360 L_t^3
+ L_h \left (36  + 648 L_t + 324 L_t^2 \right )
\Bigr ]
,
\phantom{xxx}
\label{eq:M2hpole3c}
\eeq
where the analytic forms of the decimal coefficients are
\beq
-3433.724  &\approx& -673 - \frac{17 \pi^2}{2} - 1962 \zeta_3 + 24 c_H'
,\phantom{xxxx}
\\
-2426.808 &\approx& -\frac{10491}{4} +144 \sqrt{3} \pi - 42 \pi^2 - 144 \zeta_3
,
\\
-101.016 &\approx& -\frac{855}{2} + 60 \sqrt{3} \pi
.
\eeq 
This result differs from eq.~(3.4) of ref.~\cite{Martin:2014cxa} by terms that vanish when $L_t = 0$, consistent with the approximation made in that reference. 

\begin{figure}[!t]
\begin{minipage}[]{0.545\linewidth}
  \begin{flushleft}
    \includegraphics[width=8.5cm,angle=0]{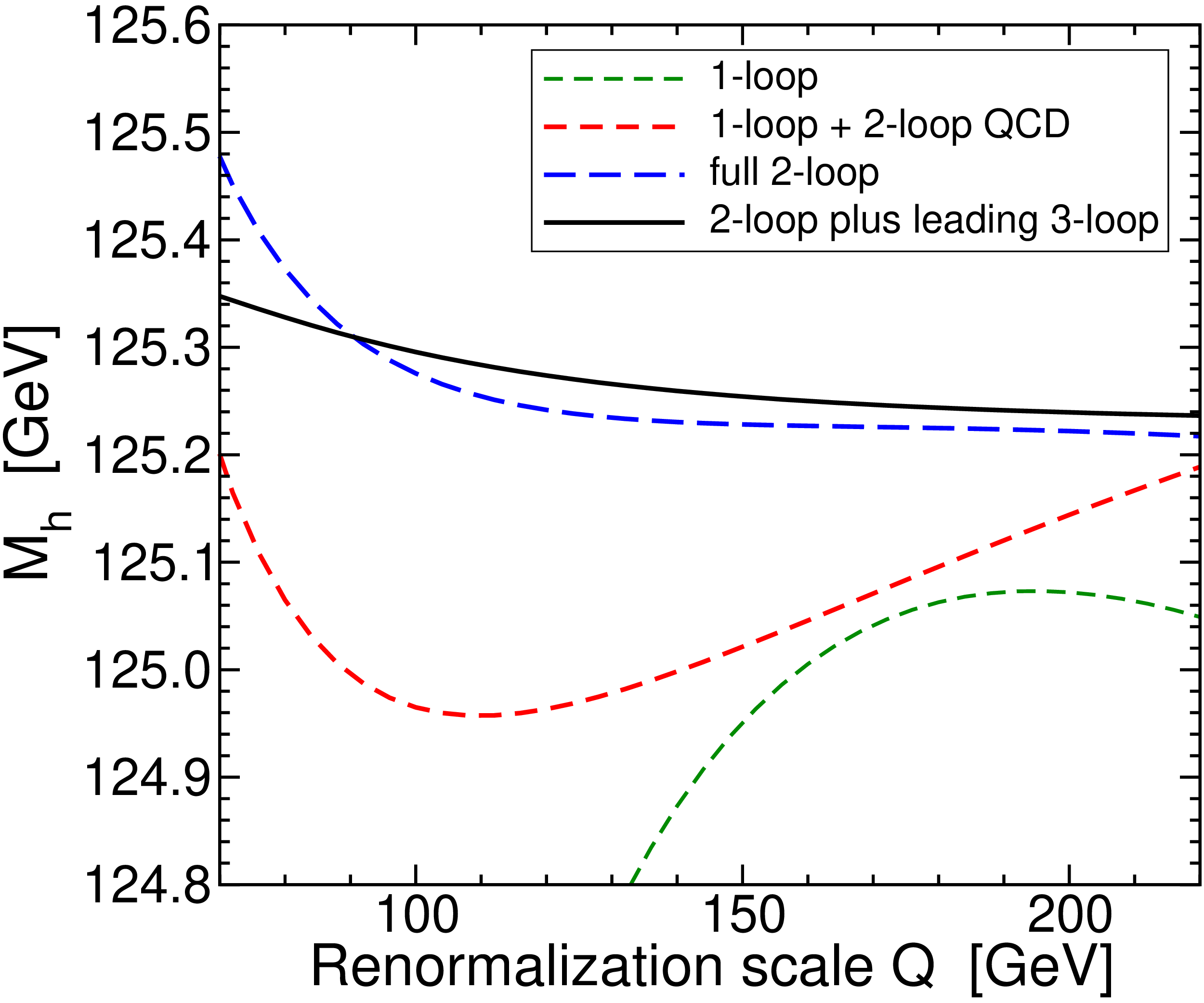}
  \end{flushleft}  
\end{minipage}
\begin{minipage}[]{0.445\linewidth}
  \caption{\label{fig:Mh}The real part of the calculated Higgs boson pole mass, as a
   function of the renormalization scale $Q$. The different lines show various 
   approximations as labeled. The \MSbar input parameters are as given in   
   eq.~(\ref{eq:referencemodelMSbar}), which provide for $M_h = 125.25$ GeV when  
   calculated at the renormalization scale $Q=160$ GeV with the best available   
   approximation as described in the text.}
\end{minipage}
\end{figure}

To recapitulate: in order to consistently include the 3-loop results given above, the non-QCD part of $\Delta_h^{(2)}$ found in the ancillary file of ref.~\cite{Martin:2014cxa} should use $s=h$ in the evaluation of the integrals, while $\Delta_h^{(1)}$ and the QCD part of $\Delta_h^{(2)}$ provided in that reference should use $s = {\rm Re}[s^h_{\rm pole}]$ determined by iteration.
All of these results for the Higgs boson pole mass have now been implemented in version 1.2 of the computer code {\tt SMDR} \cite{Martin:2019lqd}. Figure \ref{fig:Mh} shows the results for $M_h$, for the benchmark \MSbar input parameters given in eq.~(\ref{eq:referencemodelMSbar}). 
Recall that these parameters were chosen so as to give the present experimental central value from the RPP, $M_h = 125.25$ GeV, as the result of the calculation at renormalization scale $Q = 160$ GeV. The other results
in the figure were obtained by running the \MSbar parameters in eq.~(\ref{eq:referencemodelMSbar}) from the input scale
$Q_0 = 172.5$ GeV to each scale $Q$ and
re-doing the calculation. The new contributions found in this paper give the best approximation available at this writing, but still imply a scale dependence of several tens of MeV. For example, the calculated $M_h$ decreases by about 56 MeV when $Q$ is
varied from 100 GeV to 200 GeV, for fixed values of the \MSbar input parameters. This provides a lower bound on the theoretical error, and suggests that a still more refined calculation of the Higgs pole mass, to include 3-loop electroweak parts and even leading 4-loop contributions, would be worthwhile, since the experimental uncertainty on $M_h$ from future collider experiments may well be smaller \cite{deBlas:2019rxi}. It is also possible \cite{R:2021bml} to refine further the gaugeless limit by including momentum-dependent parts of the Higgs boson self-energy function.

\begin{figure}[!t]
  \begin{minipage}[]{0.59\linewidth}
    \includegraphics[width=8.8cm,angle=0]{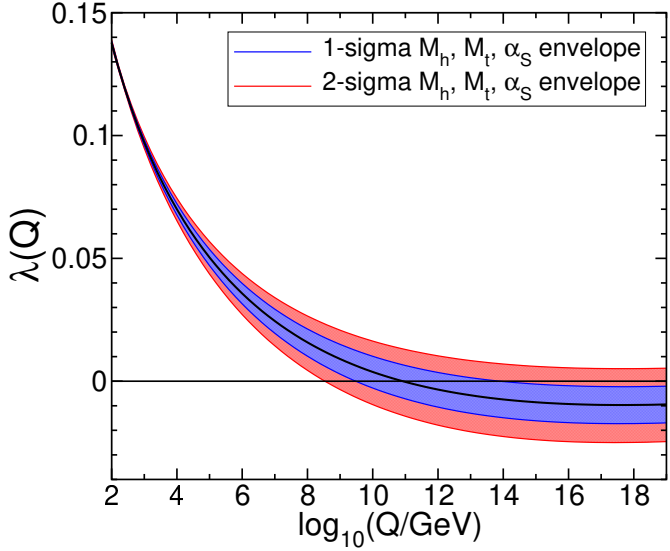}
  \end{minipage}
    \begin{minipage}[]{0.4\linewidth}
\caption{\label{fig:lambdaQ}The running Higgs self-coupling parameter $\lambda$
as a function of the \MSbar renormalization scale $Q$, using the results of this paper to relate it to $M_h$ in the most accurate available way. The central value obtained from the present experimental data as in eqs.~(\ref{eq:referencemodelOS}) and (\ref{eq:referencemodelMSbar}) is the black line. The shaded envelopes are the envelopes obtained by
varying $M_h = 125.25\pm 0.17$ GeV, $M_t = 172.5 \pm 0.7$ GeV, and $\alpha_S^{(5)}(M_Z) = 0.1179 \pm 0.0010$ in their 1-sigma and 2-sigma ranges.}
    \end{minipage}
\end{figure}

A famous feature of the observed Higgs boson mass is that the Standard Model with no extensions can then have the Higgs self-coupling $\lambda$ run negative at a scale that is far above the electroweak scale but below the Planck scale, implying a possibly metastable electroweak vacuum. This is illustrated in Figure \ref{fig:lambdaQ}, using the latest experimental values and the results of this paper to relate $M_h$ to $\lambda$ in the most accurate available way. 
As is well-known (see for example refs.~\cite{EliasMiro:2011aa} and 
\cite{Bezrukov:2012sa,Degrassi:2012ry,Buttazzo:2013uya}), the scale of possible instability is lowered
if the top-quark mass is higher, or the QCD coupling is lower, or the Higgs mass is lower, than their 
benchmark values, while it is possible for the instability to be avoided up to the Planck scale if the deviations are in the opposite directions. While improved formulas and experimental values for $M_h$ are welcome, the dominant uncertainty in these instability discussions comes from $M_t$ (or $y_t$), and the second most important uncertainty is that of $\alpha_S^{(5)}(M_Z)$, through their renormalization group running
influence on $\lambda$. 

\section{The $W$ boson pole mass\label{sec:Wboson}}
\setcounter{equation}{0}
\setcounter{figure}{0}
\setcounter{table}{0}
\setcounter{footnote}{1}

Consider the $W$-boson complex pole squared mass $s^W_{\rm pole}$ as in eq.~(\ref{eq:sXpole}). The complete 1-loop and 2-loop parts $\Delta_W^{(1)}$
and $\Delta_W^{(2)}$ were given in ref.~\cite{Martin:2015lxa}. The 3-loop QCD part splits into 8 distinct contributions with different group theory structures:
\beq
\Delta_W^{(3),g_3^4} &=&
 g_3^4 g^2 N_c C_F \biggl (
  C_G \left [\Delta_W^{(3,a)} + (n_g-1) \Delta_W^{(3,b)} \right ]
+ C_F \left [\Delta_W^{(3,c)} + (n_g-1) \Delta_W^{(3,d)} \right ]
\nonumber \\ && 
+ T_F \left [\Delta_W^{(3,e)} + (n_g-1) \Delta_W^{(3,f)} 
+ (2 n_g-1) \Delta_W^{(3,g)} + (2 n_g-1) (n_g-1) \Delta_W^{(3,h)}
\right ] \biggr )
.
\phantom{xxx}
\label{eq:DeltaW3QCD}
\eeq
The four contributions from diagrams in which the $W$ boson couples directly to 
massless quarks are relatively simple: 
\beq
\Delta_W^{(3,b)} &=& W \left (-\frac{44215}{648} + \frac{454}{9} \zeta_3 + \frac{20}{3} \zeta_5
+ \left [ \frac{41}{2} - \frac{44}{3} \zeta_3\right ] L_{-W}
- \frac{11}{6} L_{-W}^2 \right )
,
\\
\Delta_W^{(3,d)} &=& W \left (\frac{143}{18} + \frac{74}{3} \zeta_3 - 40 \zeta_5
- \frac{1}{2} L_{-W} \right )
,
\\
\Delta_W^{(3,f)} &=& \frac{8}{27}W (7t + 3W) I_{7c}(0, 0, 0, 0, 0, t, t)
+\frac{8}{243} (128 t + 43 W) I_{6c} (0, 0, 0, 0, t, t)
\nonumber \\ &&
- \frac{4}{27} (18 t + 7 W) I_{6f} (0, 0, 0, 0, t, t)
+ \frac{16}{243 t} (5 W - 17 t) I_{5c} (0, 0, 0, t, t)
+ \frac{80}{81 t} I_4 (0,0,t,t)
\nonumber \\ &&
+ \left (\frac{448}{243}\zeta_3 -\frac{4138}{729}\right ) t
+ \frac{2599}{486}{W} + \frac{40}{2187}\frac{W^2}{t} 
+ \left ( \frac{5572}{243} t - \frac{112}{9} \zeta_3 t - \frac{1660}{243} W \right )
   L_{t}
\nonumber \\ &&
-\frac{176}{81} t L_t^2
-\frac{56}{243} t L_t^3
+ \frac{2}{243} (884 t - 217 W) L_{-W}
+ \left ( \frac{80}{27} W - \frac{16}{3} t \right ) L_{t} L_{-W}
\nonumber \\ &&
+ \frac{136}{81} t L_t^2 L_{-W}
+ \frac{10}{81} W L_{-W}^2
- \frac{8}{243} (17 t + 10 W) L_{t} L_{-W}^2
,
\\
\Delta_W^{(3,h)} &=& W \left (\frac{3701}{162} - \frac{152}{9} \zeta_3 +
\left [\frac{16}{3} \zeta_3 - \frac{22}{3} \right ] L_{-W} 
+ \frac{2}{3} L_{-W}^2 
\right )
.
\eeq
In fact, $\Delta_W^{(3,b)}$, $\Delta_W^{(3,d)}$,
$\Delta_W^{(3,f)}$, and $\Delta_W^{(3,h)}$ can be obtained from, respectively,
$\Delta_Z^{(3,j)}$, $\Delta_Z^{(3,k)}$,
$\Delta_Z^{(3,l)}$, and $\Delta_Z^{(3,m)}$ in eqs.~(\ref{eq:DeltaZ3j})-(\ref{eq:DeltaZ3m}) by replacing $Z \rightarrow W$ and dividing by 2. The reason for this is that they come from exactly the same Feynman diagram topologies. 

The remaining four contributions $\Delta_W^{(3,a)}$, $\Delta_W^{(3,c)}$,
$\Delta_W^{(3,e)}$, and $\Delta_W^{(3,g)}$ in eq.~(\ref{eq:DeltaW3QCD}) are more complicated, and are relegated to an ancillary file {\tt DeltaW3}. They each have the form of
eq.~(\ref{eq:generalform}), with renormalized $\epsilon$-finite
master integrals that are a subset of eqs.~(6.2)-(6.4) of ref.~\cite{Martin:2021pnd}: 
\beq
\intI^{(1)} &=& \{\hspace{0.4pt} A(t),\> B(0,t)\hspace{0.4pt} \}
,
\\
\intI^{(2)} &=& \{\hspace{0.4pt} 
S(0,0,t),\, 
S(t,t,t),\, 
U(t,0,t,t),\, 
M(0, 0, t, t, 0)\hspace{0.4pt} \}
,
\\
\intI^{(3)} &=& \{\hspace{0.4pt} 
H(0, t, t, t, 0, t),\, 
I_4(0, t, t, t),\, 
I_{6d}(0, 0, t, 0, t, 0),\, 
I_{6d}(0, 0, t, t, 0, t),\, 
I_{6d}(t, 0, 0, 0, 0, 0),
\nonumber \\ && 
I_{6e}(0, t, 0, 0, 0, t),\, 
I_{6e}(t, 0, t, 0, 0, t),\, 
I_{6f}(0, t, t, 0, 0, t),\, 
I_{6f1}(t, 0, 0, t, 0, t),
\nonumber \\ &&
I_{7a}(0, 0, 0, 0, t, t, t),\, 
I_{7a}(0, 0, t, t, 0, 0, 0),\, 
I_{7a}(0, t, t, 0, 0, t, 0),\, 
I_{7a}(t, t, 0, 0, t, t, 0),
\nonumber \\ && 
I_{7a5}(t, t, 0, 0, t, t, 0),\, 
I_{7b}(0, 0, t, 0, t, 0, 0),\, 
I_{7c}(0, 0, t, t, 0, 0, 0),\, 
I_{7d}(0, t, 0, t, t, 0, t), 
\nonumber \\ &&
I_{7e}(0, t, t, 0, 0, 0, 0),\,
I_{8b}(0, 0, 0, t, t, 0, 0, t),\, 
I_{8c}(0, 0, 0, t, t, 0, 0, t),\, 
I_{8c}^{pk}(t, t, t, 0, 0, 0, 0, 0)
\hspace{0.4pt}\}
.
\phantom{xxxx}
\eeq
I have checked that eq.~(\ref{eq:DeltaW3QCD}) gives a pole mass $s^W_{\rm pole}$ that is renormalization group invariant through 3-loop terms of order $g_3^4$, using the
derivatives of the master integrals with respect to $Q$ found in the ancillary file
{\tt QddQ} of ref.~\cite{Martin:2021pnd}.
 
For practical numerical evaluation, after plugging in the Standard Model group theory
values in eq.~(\ref{eq:SMgrouptheory}), and applying the expansions for the master integrals in the
ref.~\cite{Martin:2021pnd} ancillary files {\tt Ioddseries} and {\tt Ievenseries} (the latter being needed only for the contribution $\Delta_W^{(3,f)}$ in which the $W$ boson couplings are to a  massless quark loop, with a top-quark loop correcting a gluon propagator), I obtain a series expansion:
\beq
\Delta_W^{(3), g_3^4} &=& g_3^4\, g^2 t \left (\delta_1^W + \delta_2^W \right ),
\eeq
where $\delta_1^W$ comes from $\Delta_W^{(3,a)}$, $\Delta_W^{(3,c)}$,
$\Delta_W^{(3,e)}$, and $\Delta_W^{(3,g)}$, which follow from diagrams where the $W$ boson couples directly to a top-bottom pair, and $\delta_2^W$ comes from
$\Delta_W^{(3,b)}$, $\Delta_W^{(3,d)}$,
$\Delta_W^{(3,f)}$, and $\Delta_W^{(3,h)}$ 
from diagrams in which the $W$ boson couples directly to light-quark pairs.
An ancillary file {\tt DeltaW3series} provided with this paper gives the
results, both analytically and numerically, to orders $\rho_W^{30}$ and $r_W^{16}$,
where
\beq
\rho_W &\equiv& \frac{W}{t} \,=\, \frac{g^2}{2 y_t^2}
\qquad
\mbox{and}\qquad
r_W = \frac{\rho_W}{4},
\eeq
and the coefficients involve $L_t = \ln(t/Q^2)$ and 
$L_{-W} = \ln(W/Q^2) - i \pi = 2 - B(0,0)\bigl |_{s = W + i \epsilon}$, as well as
$\zeta_2$, $\zeta_3$, $\zeta_4$, $\zeta_5$, $c_H'$ from eq.~(\ref{eq:definecHp}), and
\beq
c_I &=& \sqrt{3}\, {\rm Im}\left [{\rm Li}_2(e^{2\pi i/3}) \right ] \>\approx\> 1.1719536193\ldots.
\eeq
Note that $\delta_2^W$ is the same as $\delta_4^Z$ appearing in 
eqs.~(\ref{eq:DeltaZ3QCD})~(\ref{eq:delta4Z}) with the replacement $r_Z \rightarrow r_W$.
The series for $\delta_1^W$ and $\delta_2^W$ converge for $\rho_W < 1$ and $r_W < 1$, respectively, which is clearly satisfied by the relevant value of $W/t$ in the Standard Model.

The numerical form of the first few terms in the series are
\beq
\delta_1^W &=& 12.8299 + 24.9541 L_t + 63 L_t^2 - 30 L_t^3
\,+\, \rho_W (-23.800 - 54.693 L_t + 14 L_t^2) 
\nonumber \\ && 
+ \rho_W^2 (-2.327 - 17.873 L_t - 1.5 L_t^2)
\,+\, \rho_W^3 (-0.700 - 7.496 L_t - 7.2 L_t^2 ) + \ldots
,
\phantom{xxx}
\\[3pt]
\delta_2^W &=& 
r_W \left ( -56.799 - 14.758 L_{t}  - 10.667 L_t^2 + 180.381 L_{-W}
+ 21.333 L_{t} L_{-W}  - 122.667 L_{-W}^2  \right )
\nonumber \\ &&
+ r_W^2 \left [ -88.570 - 33.375 (L_t - L_{-W}) - 3.793 (L_t - L_{-W})^2 \right ]
\nonumber \\ &&
+ r_W^3 \left [ 4.074 + 2.521 (L_t - L_{-W}) + 0.406 (L_t - L_{-W})^2 \right ] + \ldots
.
\eeq
As in the case of the $Z$ boson, it is interesting to note that in this expansion in small $W/t$, the sub-leading contribution
is numerically comparable to or larger than the leading contribution (obtained by $\rho_W = r_W = 0$), depending on the choice of $Q$. This is due mostly to the term proportional to $r_W L_{-W}^2$ in the contribution from massless quark loops, because of the large magnitude of the coefficient $-122.667$ and because $L_{-W}^2 = [-i\pi + \ln(W/Q^2)]^2$ provides up to an order of magnitude enhancement.  

\begin{figure}[!t]
  \begin{minipage}[]{0.495\linewidth}
    \includegraphics[width=8.0cm,angle=0]{MWexpNEW_lo.ps}
  \end{minipage}
  \begin{minipage}[]{0.495\linewidth}
    \includegraphics[width=8.0cm,angle=0]{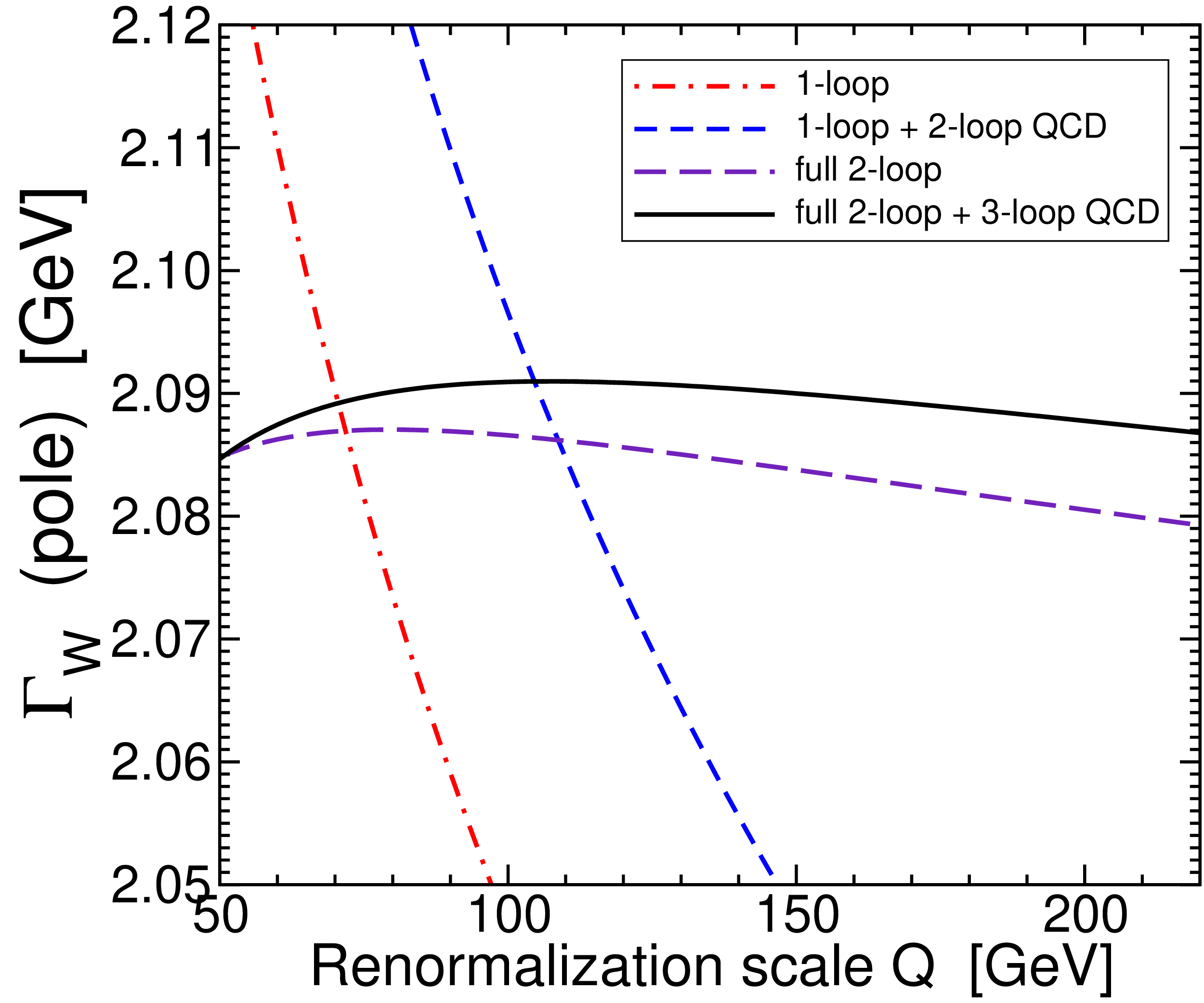}
  \end{minipage}
  \begin{center}
    \begin{minipage}[]{0.95\linewidth}
\caption{\label{fig:MW}The PDG convention $W$ boson mass, and the width $\Gamma_W$, obtained from the calculated complex pole mass $s^W_{\rm pole}$, as a function of the renormalization scale $Q$. The different lines show various approximations as labeled. The \MSbar input parameters are as given in eq.~(\ref{eq:referencemodelMSbar}).
Also shown are the experimental central values and $1\sigma$ ranges for $M_W^{\rm PDF}$ as given by the 2021 update to the 2020 RPP,
and from the 2022 result from CDF \cite{CDF:2022hxs} with statistical and systematic errors combined in quadrature.}
    \end{minipage}
  \end{center}
\end{figure}

The contribution $\Delta_W^{(3), g_3^4}$ is now implemented in the
new version 1.2 of the computer code {\tt SMDR} \cite{Martin:2019lqd}.
Figure \ref{fig:MW} shows the results for $M_W^{\rm PDG}$ and for 
$\Gamma_W$ obtained from the complex pole squared mass $s_{\rm pole}^W$, for the \MSbar input parameters
in eq.~(\ref{eq:referencemodelMSbar}) at the reference scale $Q_0 = 172.5$ GeV.
The default scale used by {\tt SMDR} v1.2 for the $W$ mass calculation is $Q=160$ GeV,
which gives $M_W^{\rm PDG} = 80.3525$ GeV and $\Gamma_W = 2.0896$ GeV. 
The results for other renormalization scales $Q$ are obtained by first running the \MSbar parameters to $Q$ 
and then re-calculating $s_{\rm pole}^W$. The three-loop QCD contribution to $M_W^{\rm PDG}$ is seen to be as large as about 6 MeV.
In the idealized case, the total $s_{\rm pole}^W$ would not depend on $Q$. The computed value of $M_W^{\rm PDG}$ varies by less than 2.4 MeV as
$Q$ is varied from 80 GeV to 180 GeV. This is significantly larger than the scale
dependence of the computed $M_Z^{\rm PDG}$ as found in Figure \ref{fig:MZ}, but compares
quite favorably to the present experimental uncertainty of 12 MeV. The range for $M_W^{\rm PDG}$ from the average of experimental data released through 2021 is
$80.379 \pm 0.012$ GeV. The CDF collaboration has recently produced a result that is substantially higher, $80.4335 \pm 0.0064_{\rm stat} \pm 0.0069_{\rm syst}$ GeV, which is in stark disagreement with the Standard Model prediction. These results are also shown in Figure \ref{fig:MW}. As seen in the right panel of Figure \ref{fig:MW}, the total variation
in $\Gamma_W$ as $Q$ varies from 60 GeV to 220 GeV is about 3.5 MeV, but the spread
is only about
2.3 MeV as $Q$ varies from 80 GeV to 180 GeV. These scale variations are improved over the full 2-loop order calculation found in
ref.~\cite{Martin:2015lxa}. For comparison, the largest parametric uncertainty contributing  to the $M_W$ prediction is that of the top-quark pole mass $M_t$.
If one fixes eqs.~(\ref{eq:referencemodelOS}) and (\ref{eq:referencemodelMSbar})  as a reference model, and then adjusts the Standard Model inputs to fit varying $M_t, M_Z, \Delta\alpha_{\rm had}^{(5)}$, and $\alpha_S^{(5)}(M_Z)$, then one finds approximately
\beq
M_W^{\rm PDG} &=& M_W^{\rm PDG, ref} 
+ \mbox{6.1 MeV} \left (\frac{M_t - M_t^{\rm ref}}{\rm GeV} \right )
+ \mbox{1.3 MeV} \left (\frac{M_Z^{\rm PDG} - M_Z^{\rm PDG, ref}}{\rm MeV} \right )
\nonumber 
\\
&&
- \mbox{1.8 MeV} \left (\frac{\Delta\alpha^{(5)}_{\rm had} - \Delta\alpha_{\rm had}^{(5), {\rm ref}}}{0.0001} \right )
- \mbox{0.7 MeV} \left (\frac{\alpha_S^{(5)}(M_Z) - \alpha_S^{(5)}(M_Z)^{\rm ref}}{0.001} \right )
\eeq  
as the prediction for the $W$ boson mass in the PDG convention, with $M_W^{\rm PDG, ref} = 80.3525$ GeV.

\begin{figure}[!t]
  \begin{minipage}[]{0.504\linewidth}
    \includegraphics[width=8.0cm,angle=0]{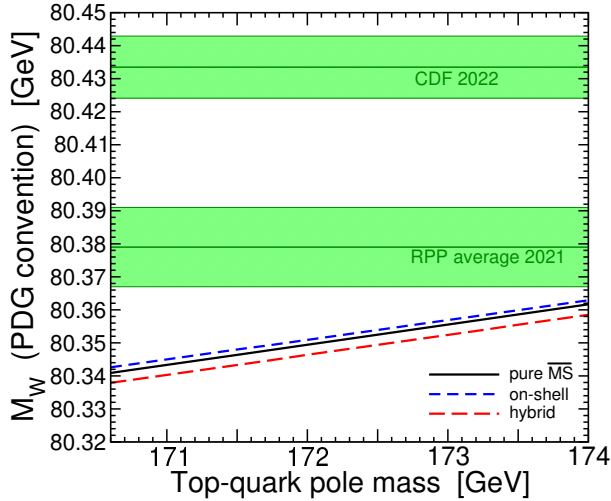}
  \end{minipage}
  \hspace{0.025\linewidth}
  \begin{minipage}[]{0.45\linewidth}
\caption{\label{fig:MWcomparison}Comparison of Standard Model predictions for the 
$W$ boson mass in the PDG convention, as a function of the top-quark pole mass $M_t$, using
data for $M_Z^{\rm PDG}$, $G_\mu$, $\alpha^{(5)}_S(M_Z)$, $\Delta \alpha_{\rm had}^{(5)}$, and $M_h$ from eq.~(\ref{eq:referencemodelOS}). The solid black line
is the pure \MSbar scheme result, obtained using {\tt SMDR} v1.2 incorporating the results of this paper.
The short dashed (blue) line is the on-shell scheme result, obtained from the interpolating
formula in ref.~\cite{Awramik:2003rn}. The long dashed (red) line is the result from the hybrid
$\overline{\rm MS}$-on-shell scheme of ref.~\cite{Degrassi:2014sxa}.  Also shown are the experimental central values and $1\sigma$ ranges for $M_W^{\rm PDF}$ as given by the 2021 update to the 2020 RPP,
and from the 2022 result from CDF \cite{CDF:2022hxs}.}
    \end{minipage}
\end{figure}

In Figure \ref{fig:MWcomparison}, I compare the prediction for $M_W^{\rm PDF}$ from {\tt SMDR} v1.2 (incorporating the results of this paper) in the pure \MSbar scheme to the corresponding results in the on-shell scheme using the interpolation formula in ref.~\cite{Awramik:2003rn}, and to those in the hybrid 
$\overline{{\rm MS}}$-on-shell scheme of ref.~\cite{Degrassi:2014sxa}, as a function of the top-quark pole mass.
The other on-shell parameters $M_Z^{\rm PDG}$, $G_\mu$, $\alpha^{(5)}_S(M_Z)$, $\Delta \alpha_{\rm had}^{(5)}$, and $M_h$ 
are chosen to be the same, and equal to the data given in 
from eq.~(\ref{eq:referencemodelOS}) from the 2021 update to the 2020 RPP, so that the results are directly comparable. (In the \MSbar scheme, this entails doing a fit to determine the Lagrangian parameters, which is readily accomplished using the C function
{\tt SMDR\_Fit\_Inputs} or the interactive command-line tool {\tt calc\_fit -int}.)
The pure \MSbar scheme gives results
between those of the on-shell and hybrid schemes, with a total spread between the three schemes
of about 4.5 MeV.

\section{Outlook\label{sec:outlook}}
\setcounter{equation}{0}
\setcounter{figure}{0}
\setcounter{table}{0}
\setcounter{footnote}{1}

In this paper, I have reported the 3-loop QCD contributions to the $W$, $Z$, and Higgs boson physical masses in the Standard Model, in the pure \MSbar renormalization scheme with a tadpole-free treatment of the Higgs VEV. The results show improved renormalization group scale independence, especially for the $W$ and $Z$ boson cases, and in all three cases the scale variation is less than the present experimental uncertainty. Alternative methods based on on-shell type schemes have already included 4-loop QCD contributions through the rho parameter, but it is not clear that these should be numerically more important than 3-loop mixed and pure electroweak contributions.
The results of this paper have all been incorporated in the latest version 1.2 of the code {\tt SMDR} \cite{Martin:2019lqd}.
Further improvements in the approach of the present paper could come from computing all of the remaining 3-loop self-energy contributions to the pole masses, which in the case of the most general diagrams will be a challenging but perhaps not insurmountable goal. 

{\it Acknowledgments:} I thank Scott Willenbrock for useful conversations regarding the parameterization of complex pole masses. This work was supported in part by the National Science Foundation grant number 2013340. 


\end{document}